# Observation of Chirality Transition of Quasiparticles at Stacking Solitons in Trilayer Graphene

Long-Jing Yin, Wen-Xiao Wang, Yu Zhang, Yang-Yang Ou, Hao-Ting Zhang, Cai-Yun Shen, and Lin He*

The Center for Advanced Quantum Studies, Department of Physics, Beijing Normal University, Beijing, 100875, China

*e-mail: helin@bnu.edu.cn

**Trilayer graphene (TLG) exhibits rich novel electronic properties and extraordinary quantum Hall phenomena owing to enhanced electronic interactions and tunable chirality of its quasiparticles. Here, we report direct observation of chirality transition of quasiparticles at stacking solitons of TLG via spatial-resolved Landau level spectroscopy. The one-dimensional stacking solitons with width of the order of 10 nm separate adjacent Bernal-stacked TLG and rhombohedral-stacked TLG. By using high field tunneling spectra of scanning tunneling microscopy, we measured Landau quantization in both the Bernal-stacked TLG and the rhombohedral-stacked TLG and, importantly, we observed evolution of quasiparticles between the chiral degree $l = 1\&2$ and $l = 3$ across the stacking domain wall solitons. Our experiment indicates that such a chirality transition occurs smoothly, accompanying the transition of the stacking orders of TLG, around the domain wall solitons. This result demonstrates the important and hitherto neglected relationship between the crystallographic stacking order and the chirality of quasiparticles in graphene systems.**

In graphene, chirality emerges naturally as a consequence of the bipartite honeycomb lattice, which creates two inequivalent Dirac cones, commonly called $K$ and $K'$, at the corners of the graphene Brillouin zone [1]. The wave functions describing the low-energy excitations in graphene are spinors and the low-energy quasiparticles have a chiral degree $l$ that depends on both the layer number and stacking order [2-5]. In graphene monolayer we have $l = 1$, and in Bernal bilayer we have $l = 2$. As a consequence, the Berry phase of quasiparticles in graphene monolayer and bilayer is $\pi$ and $2\pi$ [6-8], respectively. For trilayer graphene (TLG), there are two natural stable allotropes: one is the Bernal-stacked trilayer (or ABA trilayer); the other is the rhombohedral-stacked trilayers (or ABC trilayer) [4,9-15]. The chiral quasiparticles are quite different in the two allotropes due to the difference of the stacking orders. In the ABA trilayer, both massless ($l = 1$) and massive ($l = 2$) Dirac fermions coexist [9,10,14]; whereas the low-energy excitations in the ABC trilayer are $l = 3$ chiral quasiparticles with cubic dispersion (the corresponding Berry phase of the quasiparticles is $3\pi$) [4]. Very recently, it has been demonstrated explicitly that there are stacking domain-wall solitons separating two adjacent regions with different crystallographic stacking sequences in graphene multilayer [12,15-20]. According to the experimental observation, the transition between the ABA region and the ABC region occurs smoothly across the stacking solitons with width of the order of 10 nm, where one layer of the TLG shifts by the carbon-carbon spacing [12,15]. Such a result indicates that there should be chirality transition of quasiparticles at the stacking domain walls in the TLG. However, an experimental verification of this chirality transition is still lacking up to now.

Identifying the chiral degree of quasiparticles in graphene systems has so far been explored mainly using transport techniques [4-8], which lack spatial resolution that would be indispensable for studying the chirality transition at the stacking solitons. Our scanning probe technique allows us to achieve sub-nanometer-scale spatial resolution. In our experiment, we first used measurements based on scanning tunneling microscopy (STM) [17] to identify the stacking domain-wall solitons with different atomic structures in TLG. Then, high-field scanning tunneling spectroscopy (STS) spectra are used to determine the chiral degree of quasiparticles in both the ABA TLG and the ABC TLG. We direct observed evolution of the quasiparticles between the chiral degree $l = 1$ & $2$ in the ABA TLG and $l = 3$ in the ABC TLG across the domain walls.

In our experiment, the TLG samples were exfoliated from graphite and then the samples were transferred to both $SiO_2$/Si and graphite substrates for further studies. A facile and effective method to identify TLG with different stacking orders is Raman spectroscopy mapping [12,21,22], as shown in Fig. 1. The 2D Raman peak of the ABC TLG is more asymmetric and broader than that of the ABA TLG [Fig. 1(a)]. It is easy to visualize the spatial distribution of the stacking domains of the TLG by measuring the full-width at half-maximum (FWHM) of the 2D peak, as shown in Fig. 1(c) (see Figure S1 of Supplemental Material [23] for more experimental data). A stacking domain wall separating domains of differing stacking orders, i.e., the ABA region and the ABC region, is expected to be observed.

In order to further characterize the atomic structures and electronic properties of the stacking domain wall solitons in the TLG, we performed STM and STS measurements of the exfoliated TLG on graphite substrate (see Supplemental Material [23] for details). The STM system was an ultrahigh vacuum scanning probe microscope (USM-1500S from UNISOKU) with the magnetic fields up to 8 T. In the experiment, we first identify the exfoliated TLG on graphite substrate according to the height in the STM measurement (see Figure S2 of Supplemental Material [23] for experimental data). The exfoliated TLG decouples from the substrate when the separation between the bottom graphene sheet of the TLG and graphite is larger than the equilibrium distance 0.34 nm, or when there is a large rotation angle of the TLG with the substrate [14,17]. In this work, only the decoupled exfoliated samples are further studied. Figure 2 shows several STM images around the stacking solitons of the exfoliated TLG. We can direct discriminate the ABA region from the ABC region in the STM images recorded in low bias voltages, as shown in Fig. 2, owning to their distinct low-energy electronic structures and properties [9,10,12-15], as shown in Fig. 3(a) and 3(b). The height of the bright lines, separating the adjacent ABA and ABC regions in the topography images, depends strongly on the bias voltage used for imaging and these lines are the stacking domain wall solitons in the TLG (see Figure S3 of Supplemental Material [23] for experimental data). Besides the relatively straight stacking solitons, we also observed other different domain-wall patterns, as shown in Fig. 2(d)-(f). These different patterns provide a rich platform to explore stacking solitons with different atomic configurations.

To realize the transition of the stacking order at the domain wall, one layer of the TLG should be shifted the carbon-carbon spacing along the armchair orientation with respective to the adjacent layer (see Figure S4 of Supplemental Material [23] for schematic structures). Such a transition can be achieved through generating a localized region of either a shear strain (the change of the C-C spacing is parallel to the domain wall) or a tensile strain (the change of the C-C spacing is perpendicular to the domain wall), or something between [17,19,20]. Figure 2(b) and 2(c) show representative atomic-resolved STM images around a shear domain wall and a tensile domain wall, respectively, observed in the TLG in our STM measurements. Both of them show hexagonal-like lattices in the center of the solitons, but exhibit triangular lattices in both the adjacent ABA and ABC regions. This directly demonstrated that the studied bright lines in the STM images are the stacking domain wall solitons in the TLG. In our experiment, we observed shear-type solitons more frequently than that of the tensile-type solitons in the TLG. Similar result has also been reported in domain walls of graphene bilayer [18,19], which may arise from the fact that the energy of the shear stacking solitons is slightly lower than that of the tensile stacking solitons in graphene multilayer.

Figure 3 summarizes representative STS spectra recorded in the ABA and the ABC regions away from the stacking solitons both in the absence and in the presence of perpendicular magnetic fields. The tunneling spectrum gives direct access to the local density of states (DOS) of the surface. For the ABA TLG, we observed a typical V-shaped spectrum in zero magnetic field, as reported previously in Refs. [13,14]. For the ABC TLG, the zero field spectrum exhibits low-energy pronounced peaks, as shown in Fig. 3(c), which are generated by the flat bands around the charge neutrality point (CNP) of the ABC TLG [13,24,25]. The energy spacing ~ 10 meV of the two peaks around the CNP corresponds to an energy gap of the ABC TLG, where the substrate induces an effective interlayer bias and therefore breaks the inversion symmetry of the TLG [13].

In the presence of perpendicular magnetic fields, the spectra of the ABA TLG and the ABC TLG exhibit quite different Landau quantization, as shown in Fig. 3(d)-(i). According to the Landau levels (LLs) spectra recorded in high magnetic fields, we can deduce electronic structures and, consequently, extract the chiral degree of quasiparticles of the studied graphene systems. Figure 3(d) shows high-field spectra of the ABA TLG, which exhibit a sequence of LLs peaks of both massless Dirac fermions ($l = 1$) and massive Dirac fermions ($l = 2$). The energies of the LLs for the $l = 1$ and $l = 2$ fermions are described by [9,14,26]

$$E_n^1 = \text{sgn}(n)\sqrt{2e\hbar v_F^2 |n| B}, \qquad n = ...-2,-1,0,1,2...,$$

$$E_n^2 = \pm\sqrt{(\hbar\omega_c)^2[n(n-1)] + (U/2)^2}, \qquad n = 0,1,2...., \qquad (1)$$

respectively. Here $v_F$ is the Fermi velocity, $e$ is the electron charge, $\hbar$ is Planck's constant, $\omega_c = eB/m^*$ is the cyclotron frequency, $m^*$ is the effective mass of the quasiparticles, and $U$ is the interlayer bias. The underneath substrate can induce an interlayer asymmetric bias and therefore open a gap ($E_g \approx U$) in the parabolic bands of the ABA TLG [14]. By fitting the LL energies of the ABA TLG to Eq. (1), as shown in Fig. 3(e) and (f), we obtain $v_F = (0.976 \pm 0.005) \times 10^6$ m/s for the massless Dirac fermions ($l = 1$), and $m^* = (0.040 \pm 0.004)m_e$, $E_g \sim 12$ meV for the massive Dirac fermions ($l = 2$). This directly demonstrated that both the $l = 1$ and $l = 2$ chiral fermions coexist in the ABA TLG.

Figure 3(g) shows LLs spectra obtained in the ABC TLG (see Figure S5 of Supplemental Material [23] for more data recorded at different positions). LL spectra of quasiparticles in graphene monolayer, Bernal bilayer, twisted bilayer, and ABA TLG have been observed previously in STM experiments [14,17,26-33]. However, similar measurements for the ABC TLG have not yet to be reported, in part because of the small sample sizes of the TLG obtained in the exfoliation technique. The observed LL sequence in the ABC TLG, as shown in Fig. 3(g), is quite distinct from that reported previously in other graphene systems. We interpret this new LL sequence as the unique Landau quantization of the $l = 3$ chiral quasiparticles in the ABC TLG [4,9,34-36]. In the presence of perpendicular magnetic fields, LL spectrum of the $l = 3$ chiral Fermions in the ABC TLG was predicted to take the form [34-36]

$$E_n = E_C \pm \frac{2\hbar v_F^2 eB}{t_\perp^2}^{3/2} \sqrt{n(n-1)(n-2)}, \qquad n = 3, 4, 5..., \qquad (2)$$

where $E_C$ is the energy of the CNP, $\pm$ describes electron and hole, and $t_\perp$ is the nearest-neighbor interlayer hopping strength. For each orbital quantum number, $n$, the LLs $LL_n$ are four-fold degenerate with two from valley and two from spin. In the absence of both a external electric field and electron-electron interaction, the $n = 0$, $n$

= 1, and $n$ = 2 LLs are further degenerate and, consequently, there are 12-fold degeneracy at the CNP of the ABC TLG.

The degeneracy of the lowest LLs in the ABC TLG can be partially lifted when a bandgap is opened in the low-energy bands [4,9,37]. For example, the valley degeneracy is lifted in the presence of an interlayer potential and a finite bandgap is generated in the ABC TLG [13], as we have shown in Fig. 3(c). In such a case, the wave functions for one valley (+) of the lowest LL are mainly localized on the *A1* sites of the first layer, while the wave functions for the other valley (-) of the lowest LL are mainly localized on the *B3* sites s of the third layer [35,37]. Therefore, it is expected to observe valley-polarized Landau quantization of the $l = 3$ chiral quasiparticles since that the STM probes predominantly the DOS of the top layer. This was demonstrated explicitly in our experiment, as shown in Fig. 3(g): the magnitude of the valley-polarized LL, $LL_{(0,1,2,+)}$, is much more intense than that of the $LL_{(0,1,2,-)}$. The layer polarization of the two lowest LLs, $LL_{(0,1,2,+)}$ and $LL_{(0,1,2,-)}$, depends on the sign of electric polarity (or the sign of the energy gap) of the ABC TLG. In our experiment, we also observed opposite layer-polarized $LL_{(0,1,2,+)}$ and $LL_{(0,1,2,-)}$, (see Figure S6 of Supplemental Material [23] for more data). Therefore, the sign of the energy gap reverses across a domain wall separating two ABC domains subjecting to the opposite gate polarity and topological edge states are expected to be observed in such a domain wall, as demonstrated very recently in AB-BA domain wall of graphene bilayer [16,17].

In Fig. 3(h), we plot the energies of LLs in the ABC TLG as a function of ± $[n(n-1)(n-2)B^3]^{1/2}$. The data collapse into straight lines for both electron and hole sides indicating that the energies of LLs are scaled as $B^{3/2}$, which is a hallmark of the $l = 3$ chiral fermions. By separately using a linear fitting to the data in electron and hole branches, we obtained a bandgap of $E_g \sim 10$ meV, which is congruent with that observed in zero field [Fig. 3(c)]. In Fig. 3(i), we further plot a fan diagram of the LLs energies as a function of magnetic fields. By fitting each orbital level to Eq. (2), we obtained the interlayer coupling $t_\perp = 0.48 \pm 0.01$ eV. Similar result was also obtained in other ABC TLG (see Figure S6 as an example). Theoretically, it was predicted that $t_\perp \sim 0.5$ eV in the ABC TLG [24] and almost identical value of $t_\perp$ was extracted from transport measurement in the ABC TLG [4]. This good agreement strongly supports the analysis of our data and demonstrates that the low-energy electronic properties of the ABC TLG are governed by the $l = 3$ chiral fermions.

Finally, we carefully measured the chirality transition of the quasiparticles across the stacking solitons between the ABA and ABC TLG domains. Figure 4 shows representative high field spectra measured as a function of tip position scanning from the ABA to ABC region (see Figure S6-S8 of Supplemental Material [23] for more data). As demonstrated in Fig. 3, we can deduce the LLs of quasiparticles with different chiral degree from the high field spectra. Therefore, this measurement provides, to same extent, direct observation of chirality transition of quasiparticles across the stacking solitons of TLG. Away from the soliton, we observed LLs of the $l = 3$ chiral fermions in the ABC TLG and detected LLs of both the $l = 1$ and $l = 2$ chiral quasiparticles in the ABA TLG. Across the stacking soliton, we observed evolution of quasiparticles between the chiral degree $l = 1$&$2$ in the ABA TLG and $l = 3$ in the ABC TLG accompanying the transition of the stacking orders of TLG, as shown in Fig. 4. In the stacking domain wall soliton, LLs of both the $l = 1$&$2$ and $l = 3$ chiral fermions are detected due to spatial extension of the quasiparticles in the adjacent domains. This result unambiguously demonstrates the chirality transition of quasiparticles across the ABA-ABC domain wall soliton.

In summary, we reported different atomic structures of stacking domain wall solitions in the TLG. By using spatial-resolved LL spectroscopy, we measured distinct Landau quantization in both the ABA and ABC TLG and, more importantly, we observed the chirality transition of quasiparticles across the stacking solitons. Our result reveals a clear relationship between the crystallographic stacking order and the chirality of quasiparticles in graphene systems.

**Acknowledgments**

This work was supported by the National Basic Research Program of China (Grants Nos. 2014CB920903, 2013CBA01603), the National Natural Science Foundation of China (Grant Nos. 11422430, 11374035), the program for New Century Excellent Talents in University of the Ministry of Education of China (Grant No. NCET-13-0054), Beijing Higher Education Young Elite Teacher Project (Grant No. YETP0238). L.H. also acknowledges support from the National Program for Support of Top-notch Young Professionals.

and analysis.

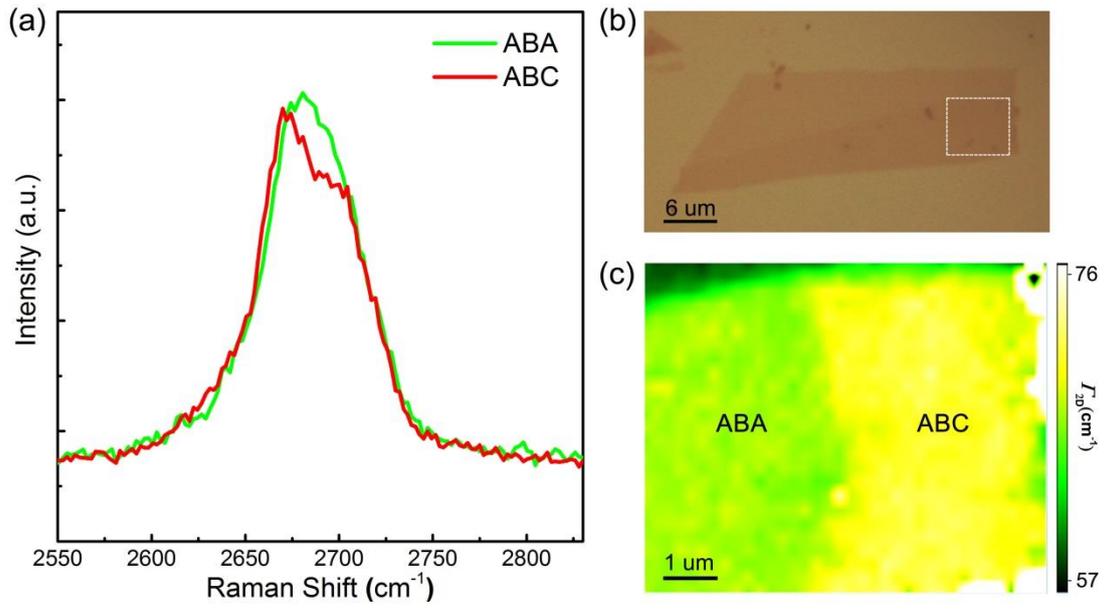

FIG. 1. Imaging the ABA and ABC TLG by Raman spectroscopy. (a) 2D Raman spectra of the ABA and ABC TLG. There is a pronounced shoulder in the 2D peak of ABC TLG. (b) Optical image of an exfoliated TLG on $SiO_2$/Si substrate. (c) Raman mapping of the region indicated by the frame in (b), showing domains of the ABA and the ABC TLG.

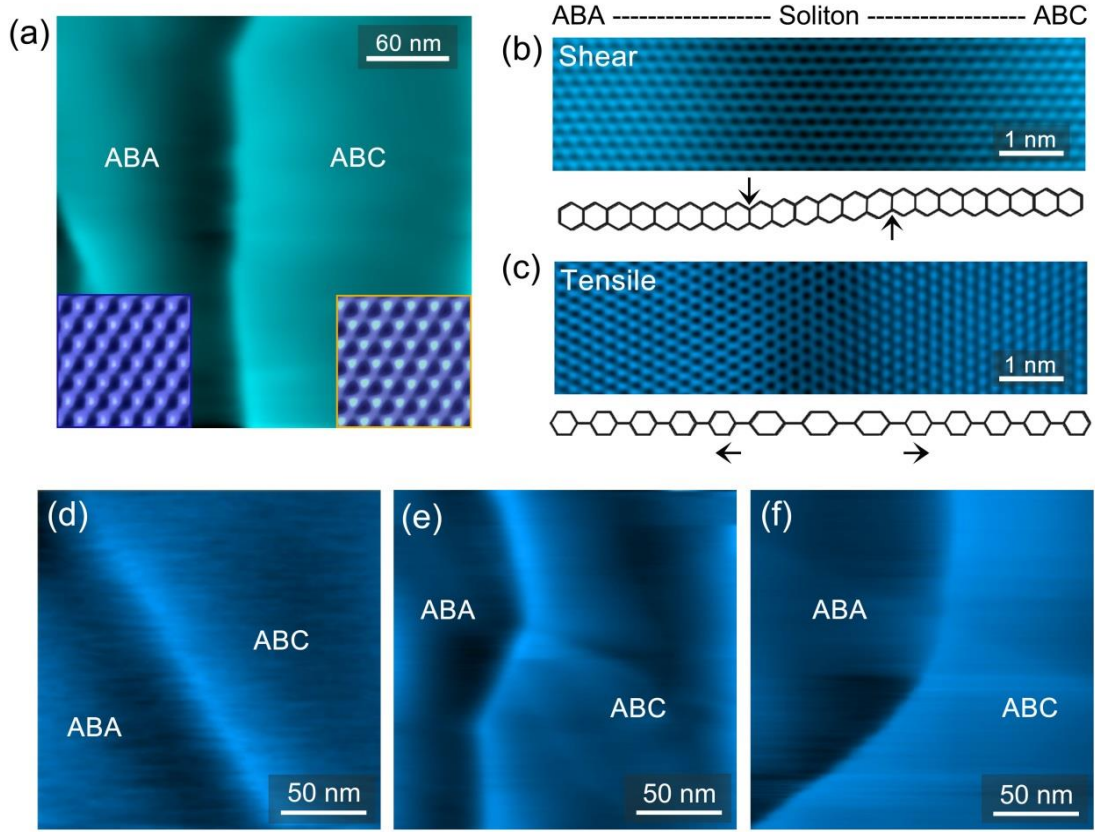

FIG. 2. Stacking domains and stacking domain-wall solitons in TLG. (a) STM topographic image ($V_b$ = 0.3 V, $I$ = 0.1 nA) of a decoupled TLG with adjacent ABA and ABC domains. Insets: atomic-resolution STM images in the ABA and ABC regions. (b and c) Atomic STM images and illustrations of shear (b) and tensile (c) stacking solitons. The black arrows indicate directions of the atomic displacement in the top graphene sheet. (d-f) Different patterns of ABA-ABC stacking solitons observed in our experiment.

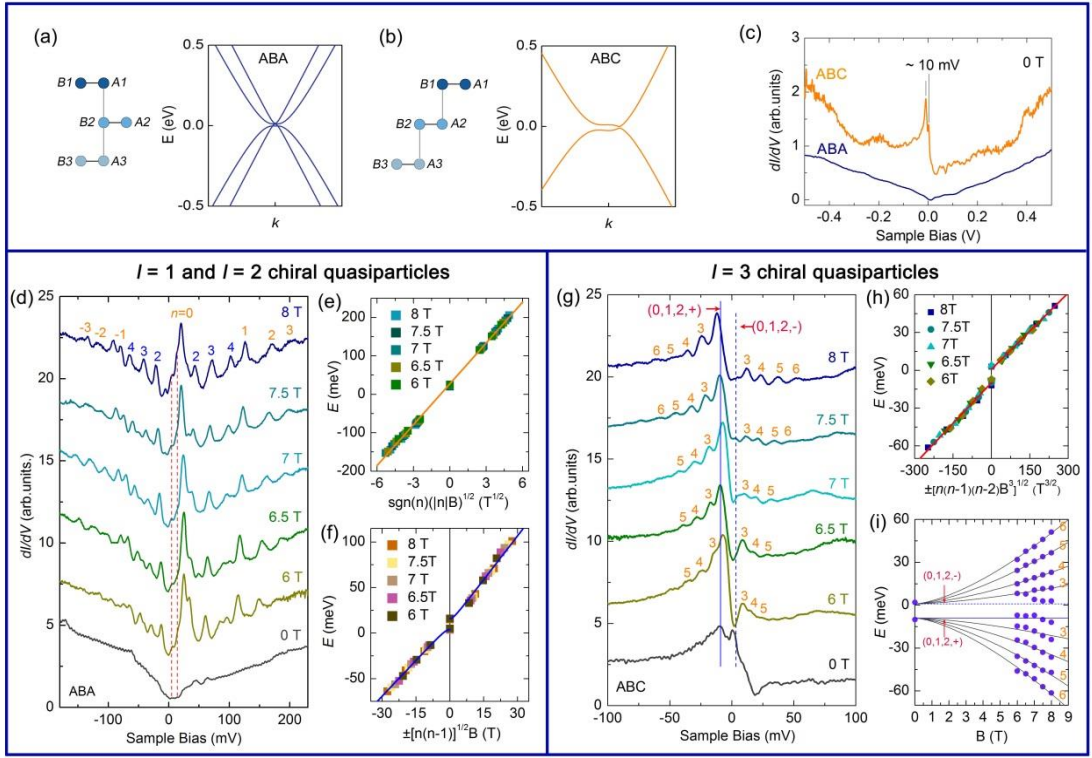

FIG. 3. Tunnneling spectra in the ABA and ABC TLG. (a and b) Schematics and band structures of the ABA (a) and the ABC (b) TLG. (c) Typical zero-field STS spectra of the ABA and ABC TLG. (d) LL spectra of the ABA TLG measured under various fields. The monolayer- and bilayer-like LLs orbital indices are marked by orange and blue numbers, respectively. The dashed lines label the gap edges in the parabolic bands. (e and f) LL peaks energies extracted from (d) showing the $B^{1/2}$ (e) and $B$ (f) dependency, respectively. The solid curves are the fitting of the data with Eq. (1). (g) STS spectra of the ABC TLG from 0 T to 8 T. LL indices are labeled. (h) LL peak energies extracted from (g) versus $\pm[n(n-1)(n-2)B^3]^{1/2}$. (i) LL peak energies versus $B$. The blue dots are the data from (g) and the black curves are fitting result with Eq. (2).

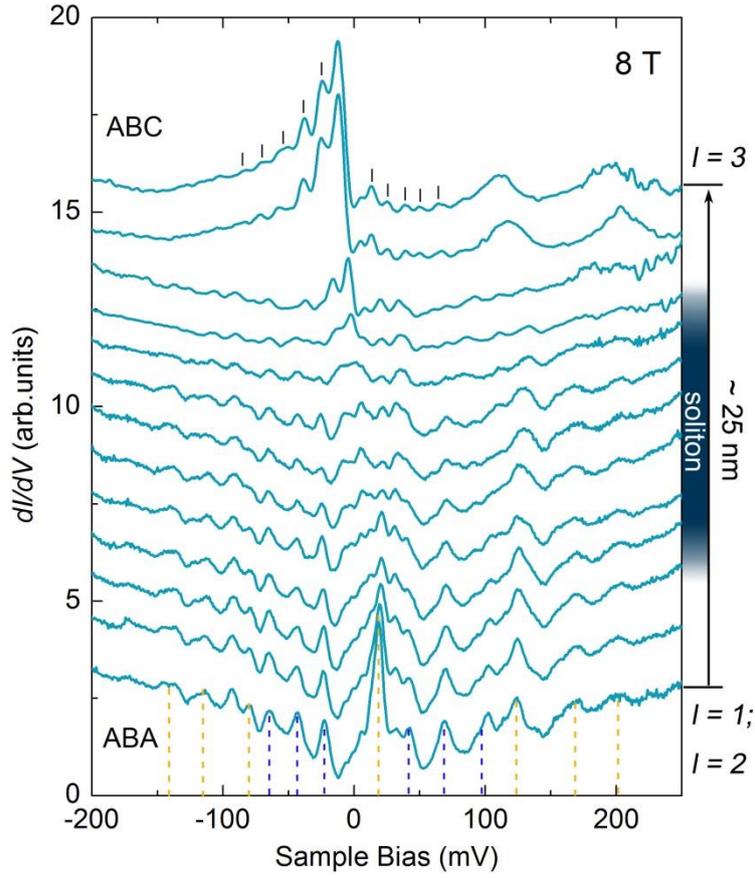

FIG. 4. Spatial evolution of the LL spectra from the $l = 1$&2 to the $l = 3$ chiral quasiparticles recorded at 8 T across an ABA-ABC shear soliton. The colored dashed lines mark the positions of LLs of the $l = 1$ and $l = 2$ chiral fermions in the ABA TLG. The black bars label the LLs of the $l = 3$ chiral fermions in the ABC TLG.